\documentclass[prl, 
twocolumn,
showpacs ,aps]{revtex4}
\usepackage{times}
\usepackage{natbib}
\usepackage{graphicx}
\usepackage{amsmath}
\usepackage{bm}
\begin{document}
\title{Vacuum-stimulated cooling of single atoms in three dimensions}
\author{Stefan Nu{\ss}mann}
\author{Karim Murr}
\author{Markus Hijlkema}
\author{Bernhard Weber}
\author{Axel Kuhn}
\author{Gerhard Rempe}
\affiliation{Max-Planck-Institut f\"ur Quantenoptik, Hans-Kopfermann-Str. 1,  D-85748 Garching, Germany.}
\date{\today}
\begin{abstract}
Taming quantum dynamical processes is the key to novel applications of quantum physics, e.g.\ in quantum information science. The control of light-matter interactions at the single-atom and single-photon level can be achieved in cavity quantum electrodynamics, in particular in the regime of strong coupling where atom and cavity form a single entity. In the optical domain, this requires a single atom at rest inside a microcavity  \cite{Ye99,McKeever03,McKeever04:2,Boca04,Maunz04,Maunz05,Mossberg91,Doherty97,Horak97,Vuletic00,Vuletic01,Domokos04,Murr03}. 
We have now discovered that an orthogonal arrangement of a cooling laser, a trapping laser and the cavity vacuum gives rise to a unique combination of friction forces that act along all three directions. This novel combination of cooling forces is applied to catch and cool a single atom in a high-finesse cavity. Very low temperatures are reached, and an average single-atom trapping time of 17 seconds is observed, which is unprecedented for a strongly coupled atom under permanent observation.
\end{abstract}


\maketitle

Cooling and trapping of single atoms in a micro-cavity  is difficult, mainly because of the limited access and the complexity of the setup. Long trapping times have been achieved for ions, but not in the small cavities required for strong coupling  \cite{Guthoerlein01,Mundt02}. Neutral atoms, in contrast, have been stored in the potential wells of a standing-wave dipole laser resonant with the micro-cavity \cite{Ye99,McKeever03,McKeever04:2,Boca04,Maunz04,Maunz05}. In these experiments, the surprisingly short trapping times originated mainly from the axial geometry of the laser-cavity system. We have now changed this geometry and use a standing-wave dipole laser oriented perpendicular to the cavity axis. Moreover, an additional pump laser induces rapid three-dimensional cooling, an effect not anticipated for a deep dipole trap. Our findings result in a deterministic strategy for assembling a permanently bound and strongly-coupled atom-cavity system. 

\begin{figure}
\centerline{\includegraphics[width=8cm]{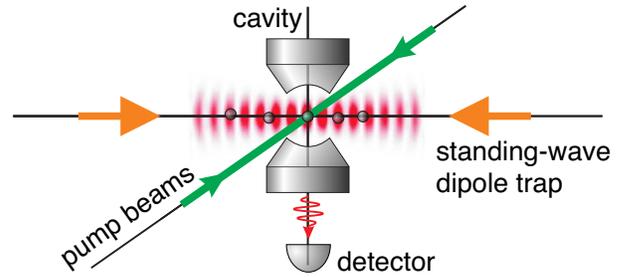}} 
\caption{\small\textbf{Schematic setup.} An atom excited by counter-propagating pump beams scatters photons into a surrounding high-finesse cavity. This goes hand-in-hand with light-induced forces that trap and cool the atom into a single potential well of a deep standing-wave dipole trap.} \label{fig:setup}
\end{figure}

As sketched in Fig.\,\ref{fig:setup}, our technique employs a standing-wave dipole-force trap and a pump beam that cross in the centre of a high-finesse optical cavity,  and run perpendicular to the cavity axis. The pump beam is near-resonant with the cavity, so that an atom in the crossing point scatters pump light into the cavity.
The momentum kicks the atom experiences when scattering these photons lead to cooling of the atomic motion along the pump and cavity directions. This process is strongly enhanced by the Purcell effect \cite{Purcell46} and has the unique advantage that cooling is effective over a large range of atomic transition frequencies. It therefore allows one to catch a free atom on its flight through the cavity and to cool it down to the bottom of a deep potential well of the dipole trap,
even though the average trap-induced AC-Stark shift of the atom increases as the localization improves. Strong cooling forces also act along the standing-wave axis and are caused by the delayed response of the atomic excitation and the intra-cavity field to a change in the atomic position. In our experiment, such a cold atom is well localized at an antinode of the standing-wave trap.

The origin of the cooling forces can be understood from a simple model based on a two-level atom. Starting from a master equation which describes the coupling of the atom to the cavity mode, the pump, and the dipole trap, one obtains four velocity-dependent forces, that we will discuss in more detail elsewhere. 
The first two cooling forces below have been theoretically predicted by Vuleti\'c et al. \cite{Vuletic00,Vuletic01} and Domokos et al. \cite{Domokos04}. We now add a dipole trap that allows for  cooling along an additional third direction. Consider an atom that is exposed to a retro-reflected pump beam of photon momenta $\pm\hbar\bm{k}_P$, with a frequency $\omega_P$ close to the resonance  $\omega_C$ of a surrounding cavity. The cavity provides a means for removing kinetic energy from the atom if the cavity resonance is blue detuned with respect to the pump $(\Delta_C=\omega_C-\omega_P>0)$. In this case, friction along the pump beams is caused by preferential absorption of photons traveling in the opposite direction than the atom. The Doppler effect shifts these photons towards the blue by $|\bm{k}_P\cdot\bm{v}|$, such that the coupled atom-cavity system becomes resonant with counter-propagating pump photons as soon as $\Delta_C+\bm{k}_P\cdot\bm{v}\approx 0$. This gives rise to a friction force,
\begin{subequations}
\label{Forces}
\begin{equation}\label{eqn:PCav}
\bm{F}_P= -4\hbar\bm{k}_P(\bm{k}_P\cdot\bm{v})
\frac{\kappa\Delta_C}{(\Delta_C^2+\kappa^2)^2} g^2P_E,
\end{equation}
along each pump beam. For a fixed low occupation probability of the atom's excited state, $P_E\simeq\Omega^2/(\Delta^2_A+\gamma^2)$ (valid for low saturation, where $2\Omega$ is the Rabi frequency of the pump, $\gamma$ is the polarisation decay rate of the atom, and $\Delta_A=\omega_A-\omega_P+\Delta_S$ is the effective pump detuning from the atomic resonance, $\omega_A$, with $\Delta_S$ being the dipole-trap induced AC-Stark shift), the friction is only determined by the cavity parameters ($g$ the atom-cavity coupling constant, $\kappa$ the field-decay rate of the cavity). 
Momentum kicks from photon emissions into the resonator lead to a similar force that acts along the cavity axis,
\begin{equation}
\bm{F}_C= -4\hbar\nabla g(\nabla g\cdot\bm{v})
\frac{\kappa\Delta_C}{(\Delta_C^2+\kappa^2)^2}P_E.
\end{equation}
Photons emitted into the direction of motion are blue detuned due to the Doppler shift. By recoil, these forward emissions also cool the atomic motion. If now the cavity is blue detuned, the emissions into the direction of motion are favored, and hence the atom is cooled along the cavity axis. This simple picture does not consider interference effects leading to a spatial modulation of the cavity field, but it holds true if the force in Eq.\,(\ref{Forces}b) is averaged over a spatial period.

The two above forces are here seen as due to distinct Doppler effects. However, they have a common origin, namely the dependency of the cavity field on the position of the atom. When the atom moves, the field takes a certain time ($\simeq\kappa^{-1}$) to adjust to a new steady-state. This time lag gives also rise to a third force
along the standing wave axis,
\begin{equation} \label{eqn:SWCav}
\bm{F}_S^{Cav}= -4\hbar\nabla \Delta_S(\nabla \Delta_S\cdot\bm{v})
\frac{\kappa\Delta_C}{(\Delta_C^2+\kappa^2)^2}
\frac{g^2P_E}{\Delta_A^2+\gamma^2}.
\end{equation}
It acts along the direction of the standing wave and depends in the same way on $\Delta_C$ as the two other friction forces. It follows that the cavity, in combination with the pump and the trap, leads to cooling in three dimensions, with forces that all have same order of magnitude.

\begin{figure}
\centerline{\includegraphics[width=8cm]{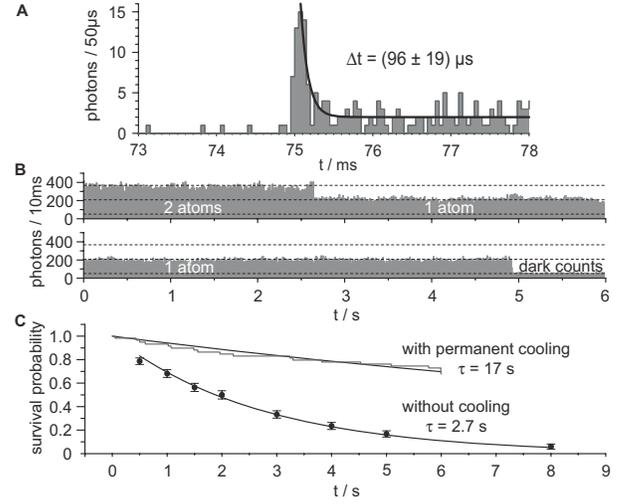}} 
\caption{\small\textbf{(A) Single-atom capture.} Photon-count rate for a pump Rabi-frequency $2\Omega=2\pi\times 50\,$MHz recorded as a function of time from the moment the standing-wave is switched on. The trace shows an atom that is captured after 75\,ms. Within 100\,$\mu$s, the scattering rate reaches a steady-state value. \textbf{(B) Atom storage.} Traces showing the photon-count rate after filtering (see text). The signal allows us to determine the atom number and the trapping time. \textbf{(C) Lifetime.} The analysis of fifty 6\,s-long traces starting with one atom yields an average lifetime of 17\,s (upper trace), while single atoms that are not exposed to the pump laser live only for 2.7\,s (lower trace). } \label{fig:TrapSteps}
\end{figure}

Our scheme also profits from a novel cooling force that is directed along the dipole trap. It can be explained by noting that an atom close to a node of the standing wave is not subject to an AC-Stark shift. For a pump frequency resonant to the atomic transition frequency $\omega_A$, the atom is resonantly pumped to the excited state. The atom now gains potential energy when moving in the standing  wave, which is then lost by spontaneous emission (rate $2\gamma$) to the ground state. This is a Sisyphus-like cooling mechanism \cite{Dalibard85,Taieb94} that uses two different fields for trapping and cooling. The resulting force,
\begin{equation} \label{eqn:SWSis}
\bm{F}_S^{Sis}= -4\hbar\nabla \Delta_S(\nabla\Delta_S\cdot\bm{v})
\frac{\Delta_A}{2\gamma(\Delta_A^2+\gamma^2)}P_E^2,
\end{equation}
\end{subequations}
is cavity-independent and provides cooling also if $\Delta_C<0$. As shown below, this force alone is sufficient to increase the trapping time. However, for $\Delta_C >0$, cavity forces dominate, and permanent photon scattering into the cavity takes place. Apart from cooling, all forces fluctuate and lead to heating of the atomic motion. The heating rate follows the Lorentzian-shaped cavity resonance (apart from Eq.\,(\ref{Forces}d)). In analogy to free-space laser cooling \cite{Cohen92}, a cavity-Doppler limited temperature around $k_BT\simeq\hbar\kappa$ is expected for $\Delta_C\simeq\kappa$. This unique combination of friction forces is unprecedented, and it allows one to cool single dipole-trapped atoms in a cavity along all three directions. We here apply this novel combination of cooling forces to catch and cool a single atom in a high-finesse cavity, as discussed in the following. 

In the experiment, we use a dipole-force trap to guide $^{85}$Rb-atoms over a distance of 14\,mm from a magneto-optical trap (MOT) into the cavity. This trap is formed by a single horizontally running beam of an Yb:YAG laser, with focus between MOT and cavity. Once the atoms reach the cavity, we switch to a standing-wave dipole trap, formed by a pair of counter-propagating Yb:YAG laser beams (2\,Watt, 1030\,nm, waist w$_0 =16\,\mu$m), tightly focused  in the centre of the cavity. The antinodes of the standing wave represent 2.5\,mK deep potential wells, i.e.\ an AC Stark shift of the atomic transition frequency in the centre of the wells of $\Delta_S=2\pi\times 100\,$MHz.

The 0.5\,mm long cavity is formed by two mirrors of 5\,cm radius of curvature that have different transmission coefficients (2\,ppm and 95\,ppm). The relevant atom-cavity parameters are $(g_0, \kappa, \gamma)\,=\,2\pi\times(5, 5, 3)\,$MHz, where $g_0$ is the atom-cavity coupling constant in an antinode, averaged over all magnetic sublevels of the $5^2S_{1/2}(F=3)$ to $5^2P_{3/2}(F'=4)$ transition. A Pound-Drever-Hall technique is used to lock the frequency of the TEM$_{00}$ mode to the atomic transition frequency. For this purpose, we use a reference laser that is red detuned  by eight free spectral ranges (5\,nm) from the atomic resonance. This laser acts as an additional standing-wave dipole trap along the cavity axis. Its potential wells are about 30\,$\mu$K deep and show a good overlap with the resonant mode in the centre of the cavity. Together with the Yb:YAG standing-wave trap, it forms a 2D optical lattice (calculated trap frequencies: $\nu_{sw}\simeq670$\,kHz in direction of the standing-wave trap, $\nu_{cav}\simeq$\,100\,kHz along the cavity axis, and $\nu_\perp\simeq10$\,kHz orthogonal to the cavity and the trapping laser).

\begin{figure}
\centerline{\includegraphics[width=8cm]{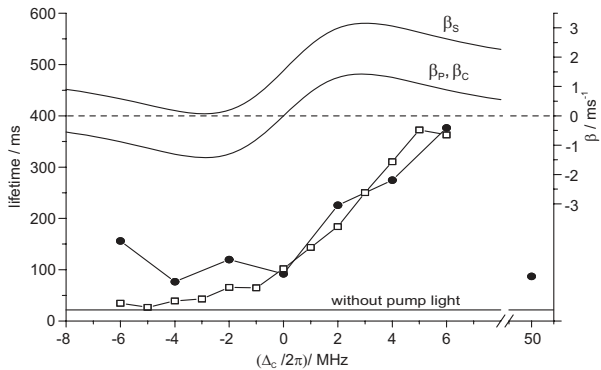}} 
\caption{\small\textbf{Trapping time and friction forces} as a function of the cavity detuning, $\Delta_C$, with either $\omega_P=\omega_A$ (circle), or $\omega_C=\omega_A$ (square). The data points show the lifetime of atoms that are subject to strong parametric heating. An extended trapping time is found for almost all detunings if pump light is present. A maximum is reached when the cavity is slightly blue detuned with respect to the laser. This is also expected from the friction coefficients $\beta=-F/m v$ from Eqs.\,(\ref{Forces}). Although the forces along pump and cavity axis change sign, the total friction along the standing-wave dipole trap, $\beta_S=\beta_S^{cav}+\beta_S^{Sis}$, is always positive due to the Sisyphus effect.}
\label{fig:DetunedCool}
\end{figure}

In addition to these two conservative dipole traps, we continuously drive the $F=3$ to $F'=4$ transition with a pump laser that runs orthogonal to the cavity axis, at an angle of 45$^\circ$ to the standing-wave trap (Fig.\,\ref{fig:setup}). Together with a repump laser $(F=2 \rightarrow F'=3)$, the beam has a focus $(w_0=35\,\mu$m$)$ at the intersection of cavity mode and dipole trap, and is retroreflected to balance its radiation pressure. To avoid an intensity modulation of the pump, the two counter-propagating beams have orthogonal polarization. Averaged over all magnetic sub-levels, it drives the atoms with a Rabi frequency $2\Omega=2\pi\times 30\,$MHz  (all data, except Fig.\,\ref{fig:TrapSteps}A). Under these circumstances, the Purcell effect gives rise to photon scattering into the cavity at a rate
\begin{equation} \label{eqn:Rate}
R_{scat}\simeq 2\kappa\cdot\frac{g^2}{\Delta_C^2+\kappa^2}\cdot\frac{\Omega^2}{\Delta_A^2+\gamma^2}
\end{equation}
for a single atom. For $\Delta_C=0$, this gives a scattering rate of $R_{scat}\simeq 1400/$ms.  In the experiment, we must take into account that the atom stops fluorescing whenever it falls into the dark state $5^2S_{1/2}(F=2)$, and that repumping to $F=3$ takes some time due to the large AC-Stark shift. This leads to blinking, and from the measured count rate, we estimate that photons are scattered at the above rate only 1/5 of the time. Furthermore, due to losses in the imaging system and the limited quantum efficiency of the detector, only 5\% of the photons that are scattered into the cavity mode are finally detected behind the cavity mirror of  higher transmission.

Once the atoms have been brought into the vicinity of the cavity, they are randomly distributed over the potential wells of the standing wave, with a small probability that an atom actually sits in the cavity. Some atoms have enough kinetic energy to move from well to well, and as soon as such an atom enters the intersection of cavity mode and pump laser $(\Delta_C/2\pi=+2\,$MHz and $\omega_P=\omega_A$ for capturing$)$, it scatters photons into the cavity and is therefore cooled. In Fig.\,\ref{fig:TrapSteps}A, an experimental trace is shown where a single atom suddenly appears in the cavity. First the photon scattering rate is high, as the initially hot atom is poorly localized in the trap and the average experienced Stark shift $\Delta_S$ is small. This changes as the atom gets colder and therefore is better localized in the potential well. After $\Delta t\simeq 100\,\mu$s, the atom reaches its final temperature and scatters at a much lower (but constant) rate, since it now experiences a much higher $\Delta_S$ close to the bottom of the potential. Starting from the simple estimation that the total kinetic energy, $E$, is lost during $\Delta t$ with $\dot{E}\simeq E/\Delta t$, one calculates a mean friction coefficient, $\beta=\dot{E}/2E\simeq 1/2\Delta t$, between $5/$ms (raw data) and $25/$ms (assuming blinking as discussed above). This is reasonably close to the expected value, $\beta = 14/$ms, that one obtains from Eqs.\,(\ref{eqn:SWCav}) and (\ref{eqn:SWSis}) with $\bm{F}=-\beta\, m\bm{v}$ ($m$ the atomic mass).

To prevent further atoms from penetrating into the cavity mode, we apply a filtering procedure. This is accomplished by a 10\,ms long interruption of the standing-wave trap. During this time, atoms inside the cavity stay trapped in the shallow intra-cavity dipole trap, but the other atoms are lost. If we turn the trap off and back on adiabatically,  the probability for a caught atom to survive this procedure is higher than 50\%. As shown in Fig.\,\ref{fig:TrapSteps}B, we then measure the photon rate to determine the exact number of atoms in the cavity \cite{McKeever04:2}. The signal shows only small variations, which indicate a nearly constant atom-cavity coupling, and pronounced individual steps, that occur whenever an atom is lost.

As shown in Fig.\,\ref{fig:TrapSteps}C, dipole-trapped atoms show an average lifetime of 2.7\,s if we switch off the pump laser. However, if an atom is continuously illuminated and coupled to the cavity with $\Delta_C\geq0$, the lifetime increases, reaching values exceeding 20\,s. This impressively demonstrates that a strong cooling mechanism is active. For $\Delta_C=0$, in particular, a lifetime of about 17\,s is measured. In this case, no cavity cooling is expected, and within our model, the long lifetime comes solely from the Sisyphus-like mechanism, Eq.\,(\ref{eqn:SWSis}). Cavity cooling, Eqs.\,(\ref{eqn:PCav}--\ref{eqn:SWCav}), is expected to give much longer trapping times for $\Delta_C\approx +\kappa$. For technical reasons, however, longer trapping times could not be registered in our experiment.  We have therefore deliberately modulated the depth of the trapping potential, i.e. the intensity of the Yb:YAG trapping laser, by 30\% at a frequency of 7\,kHz. This leads to parametric heating and shortens the trapping time, so that systematic lifetime measurements can be made within a reasonable time. Without cooling, the modulation reduces the trapping time to $(22\,\pm\,5)\,$ms. We now have performed measurements such as those in Fig.\,\ref{fig:TrapSteps}C for several different cavity detunings. The single-atom lifetime as a function of $\Delta_C$ is plotted in Fig.\,\ref{fig:DetunedCool}. Obviously, the lifetime increases to about 100\,ms as soon as the pump laser is present, even if $\Delta_C$ is so large ($2\pi\times 50\,$MHz) that the cavity has no effect. Moreover, a cavity close to resonance significantly extends the lifetime. For a blue detuned cavity, with $\Delta_C \simeq +\kappa$, a 20-fold increase of the lifetime to a maximum of $\simeq 400\,$ms is obtained, whereas a red detuned cavity, with $\Delta_C\simeq-\kappa$, leads to a minimum lifetime. To compare these results to the theoretical model, the expected friction coefficients are also shown in Fig.\,\ref{fig:DetunedCool}. Although these values cannot be directly compared with the lifetime, qualitative conclusions can be drawn. First, the cavity-independent Sisyphus cooling, predicted only along the standing-wave axis, accounts for the cavity-independent increase of the lifetime with respect to atoms left in the dark. The variation of the lifetime with $\Delta_C$ finds its correspondence in the predicted course of the friction coefficients due to the cavity cooling for all three dimensions. Obviously,  cavity cooling increases the trapping time by a factor of four if compared to Sisyphus cooling alone. Without further limitations, such as background gas collisions, we can therefore expect that the 17\,s--long lifetime observed without the trap modulation at $\Delta_C=0$ increases to about one minute in the presence of the cavity cooling at $\Delta_C\approx+\kappa$. 

\begin{figure}
\centerline{\includegraphics[width=8cm]{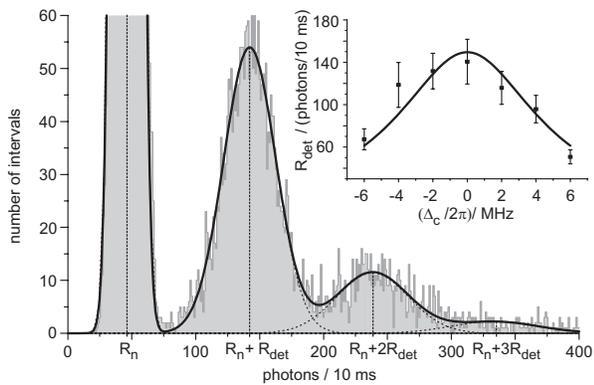}} 
\caption{\small\textbf{Photon-count histogram} measured for $\Delta_C/2\pi=+4\,$MHz using 10\,ms-long count intervals. The maxima (left to right) stem from  background noise (rate $R_n$) and from one, two or more atoms. The solid line is a fit of four Gaussians to the data. They are centered at $R_n+n\cdot R_{det}$, and are $\left(R_n+n\cdot R_{det}+ n\cdot \sigma_{R_{det}}^2\right)^{1/2}$ wide  $(n = 0, 1, 2, 3)$. $R_{det}$ is the detected photon-count rate per atom, which is also shown in the inset as a function of $\Delta_C$, together with the expectation from  Eq.\,(\ref{eqn:Rate}) as a solid line, with adapted overall amplitude.}
\label{fig:Histogram}
\end{figure}

For the same data, we have also analyzed the average count rate per atom using 10\,ms-long time bins. Fig.\,\ref{fig:Histogram} depicts the photon-count histogram. The well-distinct peaks stem from dark counts and traces with one, two or more trapped atoms. From a fit to these data, we can derive the average count rate per atom, $R_{det}$, and its statistical spread, $\sigma(R_{det})$, that is corrected for shot noise. The results are plotted in the inset of Fig.\,\ref{fig:Histogram} (with the errorbars indicating $\sigma(R_{det})$) as a function of $\Delta_C$. If we assume that all variations of $R_{det}$ are caused by variations in the atom-cavity coupling, then we obtain from Eq.\,(\ref{eqn:Rate}) $\Delta g/g=\sigma(R_{det})/2 R_{det}=\pm 8.6$\% (for $\Delta_C/2\pi=+4\,$MHz). This can only be explained with an atom distribution among the different wells of the standing-wave that is less than $\pm 9\,\mu$m wide. Under the assumption that the distribution in the filter phase is mapped to the standing-wave trap during the adiabatic transfer, this indicates a temperature below $6\,\mu$K during filtering.

To get an estimate of the temperature in the deep 2D-lattice, we analyzed the auto-correlation function of the emitted photon stream. In this signal, a periodic modulation at  2$\times\nu_{cav}$ (the trap frequency along the cavity axis) is found, with a visibility of about 5\,\%. This is indeed expected for an atom that oscillates in the weak intra-cavity dipole trap, without ever reaching the nodes of the cavity mode. The $k_B\times 30\,\mu$K depth of the intra-cavity trap can therefore be seen as an upper limit of the atomic energy. The mean kinetic energy cannot surpass 50\,\% of this value, which corresponds to a temperature of $15\,\mu$K. Assuming the same temperature in all directions, this limits the mean vibrational quantum number for the motion along the standing-wave dipole trap to $\bar n =0.13$, i.e.\ the atom is in the vibrational ground state with at least 88\,\% probability.

Our experiment demonstrates how to capture single atoms with a three-dimensional cavity cooling scheme. Qualitatively, our results show good agreement with the predictions of a model valid for a two-level atom, although an understanding of the remarkably good localization of the atom calls for a more detailed theoretical analysis, including polarization effects of the incident pump laser \cite{Cirac93:3}, the multilevel structure and even the quantum motion of the atom \cite{Cirac95:2,Zippilli05}. At present, our method allows the preparation of an exactly known number of atoms at low temperatures in the centre of a high-finesse optical cavity. With average trapping times exceeding 15 seconds, single-atom cavity physics is now at a stage where one can start fully controlled atom-photon experiments in the strong atom-cavity coupling regime.

\begin{acknowledgements} 
This work was supported by the Deutsche Forschungsgemeinschaft (SPP 1078 and SFB 631) and the European Union [IST (QGATES) and IHP (CONQUEST) programs]. We are also grateful to Simon Webster for helpful comments.
\end{acknowledgements}

\end{document}